\documentstyle[preprint,tighten,aps,psfig]{revtex}

\newcommand{\order}{{\cal O}}
\newcommand{\drm}{{\mathrm d}}

\begin{document}

\preprint{BNL-HET-99/20, hep-ph/9908439}

\title{\boldmath Muonium Decay}

\author{Andrzej Czarnecki}
\address{Physics Department, Brookhaven National Laboratory,
Upton, NY 11973}

\author{G.~Peter Lepage}
\address{Newman Laboratory of Nuclear Studies,
Cornell University, Ithaca, NY 14853-5001}

\author{William J. Marciano}
\address{Physics Department, Brookhaven National Laboratory,
Upton, NY 11973}

\maketitle

\begin{abstract}
Modifications of the $\mu^+$ lifetime in matter due to muonium
($M=\mu^+e^-$) formation and other medium effects are
examined. Muonium and free $\mu^+$ decay spectra are found to differ
at ${\cal O}(\alpha m_e/m_\mu)$ from Doppler broadening and  ${\cal
O}(\alpha^2 m_e/m_\mu)$ from the Coulomb bound state potential.
However, both types of corrections are shown to cancel in the total
decay rate due to Lorentz and gauge invariance respectively, leaving a
very small time dilation lifetime difference, $(\tau_M -
\tau_{\mu^+})/\tau_{\mu^+} = {\alpha^2 m_e^2\over 2m_\mu^2} \simeq
6\times 10^{-10}$, as the dominant bound state effect.  It is argued
that other medium effects on the stopped $\mu^+$ lifetime are
similarly suppressed.
\end{abstract}

\vspace*{0.2cm} 
The muon lifetime, $\tau_\mu$, is very well measured.  Its current
world average \cite{PDG98}
\begin{eqnarray}
\tau_\mu = 2.197035(40)\times 10^{-6}\; \mbox{sec}
\label{eq1}
\end{eqnarray}
exhibits an uncertainty of only 18 ppm.  From that lifetime, the Fermi
constant (denoted here by $G_\mu$) is determined via the defining
relationship \cite{berman62}
\begin{eqnarray}
\tau_\mu^{-1} &=& \Gamma(\mu\to \mbox{all}) 
= {G_\mu^2 m_\mu^5\over 192\pi^3}
f\left( {m_e^2\over m_\mu^2} \right)
\left( 1+{3\over 5}{m_\mu^2\over m_W^2} \right)
(1+{\rm R.C.}),
\nonumber \\
f(x) &\equiv& 1-8x +8x^3 -x^4-12x^2\ln x,
\label{eq2}
\end{eqnarray}
where $m_\mu$ is the muon pole mass and
R.C. stands for QED radiative corrections to muon decay as
calculated in an effective local V-A theory.  Other standard model
electroweak loop corrections as well as possible ``new physics''
effects are absorbed in $G_\mu$.  

The R.C. in (\ref{eq2}) have been computed 
\cite{berman62,vanRitbergen:1998yd}
through ${\cal O}(\alpha^2)$ and higher order logs have been obtained
using the renormalization group \cite{roos71}.  Altogether, one finds 
\cite{Marciano:1999ih,Nir:1989rm}
\begin{eqnarray}
{\rm R.C.} &=& {\alpha\over 2\pi}
\left[
{25\over 4} - \pi^2 + {m_e^2\over m_\mu^2} 
 \left( 48\ln {m_\mu\over m_e} - 18 - 8\pi^2\right) 
\right]
\nonumber \\
&& \quad \times \left[
 1+ {\alpha\over \pi}\left( {2\over 3}\ln {m_\mu\over m_e}-3.7\right)
+ {\alpha^2\over \pi^2} \left( {4\over 9}\ln^2 {m_\mu\over m_e}
-2.0 \ln {m_\mu\over m_e}+C\right)+\ldots\right]
\nonumber \\
\alpha^{-1} &=& 137.03599959(40)
\label{eq3}
\end{eqnarray}
where $C$ corresponds to unknown non-logarithmic ${\cal
O}(\alpha^3/\pi^3)\sim 10^{-8}$ corrections which are assumed to be
insignificant.  Employing (\ref{eq1},\ref{eq2},\ref{eq3}) leads to 
\begin{eqnarray}
G_\mu = 1.16637(1)\times 10^{-5}\; \mbox{GeV}^{-2}.
\end{eqnarray}

The Fermi constant can be compared with other precise measurements
such as $\alpha$, $m_Z$, $\sin^2\theta_W$, $m_W$, etc., and used to
test the consistency of the Standard Model at the quantum loop level.
For example, $G_\mu = \pi\alpha/ \sqrt{2}m_W^2(1-m_W^2/m_Z^2)(1-\Delta
r)$ where $\Delta r\simeq 0.0358$ (for $m_{\rm top}=174.3$ GeV and
Higgs mass $m_H\simeq 125$ GeV) represents calculable electroweak
radiative corrections \cite{frame}.  In that role, $G_\mu$ helped
predict the top quark's mass before its discovery and currently
constrains the Higgs mass to relatively low preferred values $m_H
\lower.7ex\hbox{$
 \;\stackrel{\textstyle<}{\sim}\;$}
220$ GeV.  In addition, it provides a sensitive probe of ``new
physics'' such as SUSY, Technicolor, Extra Dimensions
etc. \cite{Marciano:1999ih}.

Recently, there have been several proposals
\cite{carey,kirkby,nagamine} to further improve the measurement of
$\tau_\mu$ (and thereby $G_\mu$) by as much as a factor of 20,
bringing its uncertainty down to an incredible $\pm 1$ ppm.  Of
course, such an improvement can only by fully utilized if the other
electroweak parameters with which it is compared reach a similar level
of precision and radiative corrections to their relationship
(e.g. $\Delta r$) are computed at least through 2 loops.  That
confluence of advances appears unlikely in the foreseeable future.
Nevertheless, given the fundamental nature and importance of $G_\mu$,
efforts to improve its determination should be strongly encouraged and
pushed as far as possible.  In that spirit, we examine in this paper
several theoretical concerns that must be addressed in any $\pm 1$ ppm
study of $\tau_\mu$.

Precision muon lifetime experiments involve stopping $\mu^+$ in
material.  At some level, the medium in which it comes to rest will
affect the muon decay rate.  The most straightforward issue to
consider is the formation of muonium ($M=\mu^+e^-$) and its bound
state effect on $\tau_\mu$.

During the slowing down of $\mu^+$ in matter, muonium will form and be
ionized many times \cite{Brewer,Seeger}.  Eventually it comes to
(thermal) rest at which point the muon or muonium atom has a room
temperature kinetic energy $\sim 0.04$ eV or thermal velocity
$\beta^{\rm thermal}_\mu \simeq 2.7\times 10^{-5}$.

  The actual fraction of stopped $\mu^+$
that end up as muonium and decay while in that bound state is very medium
dependent.  It can range from a small fraction in metals to nearly
100\% in some materials.  If a significant modification of $\tau_\mu$
occurred in muonium, a correction would have to be applied.  For
example, a published study \cite{Chatterjee:1992yi} has claimed a
0.999516 reduction factor for the bound state decay rate.  Such a
large $-484$ ppm shift would be difficult to correct for at the $\pm
1$ ppm level unless the muonium formation fraction was very precisely
known.  It would also impact the interpretation of existing $\tau_\mu$
measurements \cite{Hertzog} and might indicate the possibility of
other large medium effects.

Given its potential importance, we have reexamined the muonium bound
state effect on the muon lifetime.  As we shall show, the leading
correction turns out to be remarkably
small, ${\alpha^2 m_e^2 \over 2m_\mu^2} \simeq 6\times 10^{-10}$ and
easy to calculate.

We begin by recalling some basic properties of muonium \cite{muoniu}.
It is a 
$\mu^+e^-$ Coulombic bound state with hydrogenic features.  The
reduced mass
\begin{eqnarray} 
m={m_em_\mu\over m_\mu+m_e} \simeq 0.995 m_e
\end{eqnarray}
is slightly below the electron mass, but the difference is
inconsequential for the considerations here and can be neglected.  The
energy levels are given by  
\begin{eqnarray}
E_n = -{\alpha^2m\over 2n^2} = -13.5 \, {\rm eV}/n^2, \qquad
n=1,2\ldots
\label{eq6}
\end{eqnarray}
For the $n=1$ ground state the virial theorem tells us that the
average (electron) kinetic energy is
\begin{eqnarray}
\langle T\rangle = -E_1 = 13.5 \, {\rm eV}
\end{eqnarray}
while the average bound state potential energy is 
\begin{eqnarray}
\langle V\rangle = E_1 - \langle T\rangle =\alpha^2 m  = -27.0 \, {\rm
eV}. 
\label{eq8}
\end{eqnarray}
Also, the $n=1$ momentum distribution is given by  \cite{bethe77}
\begin{eqnarray}
4\pi p^2 \left| \psi(p) \right|^2 = {32\over \pi}\alpha m 
\left( {\alpha m \over p^2 + \alpha^2 m^2} \right)^4 p^2.
\label{eq9}
\end{eqnarray}
It peaks at $p^2 =\alpha^2m^2 /3$ which is somewhat below the average
$\langle p^2 \rangle = \alpha^2 m^2$.  
In that configuration the $e^-$ and $\mu^+$ have equal but opposite
momenta and r.m.s.~velocities, $\beta_{\rm r.m.s.}\equiv \langle
\beta^2\rangle ^{1/2}$, $\beta_e = \alpha \simeq 1/137$, and
$\beta_\mu \simeq \alpha m_e / m_\mu \simeq 3.5\times 10^{-5}$.  Note that
the muon's r.m.s. bound state velocity is similar (somewhat larger) in
magnitude to its room temperature thermal velocity.

\begin{figure} 
\begin{minipage}{16.cm}
\[
\hspace*{-15mm}
\mbox{ 
\psfig{figure=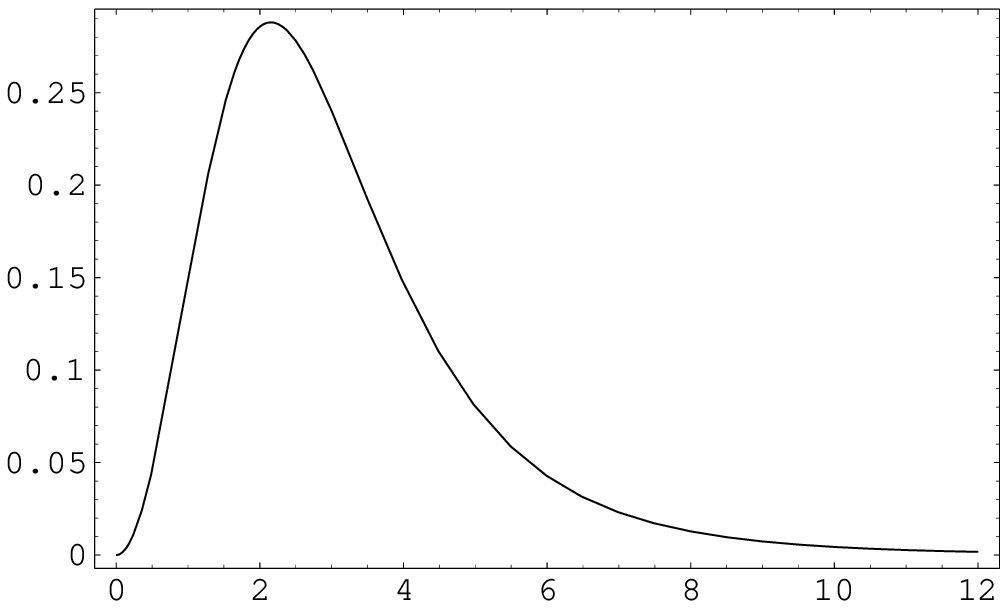,width=100mm}
}
\]
\end{minipage}
\begin{picture}(0,0)(0,0)
\put(380,15) {p [keV]}
\put(20,170) {$4\pi p^2 |\psi(p)|^2$}
\end{picture}

\caption{Momentum probability distribution  in muonium
  $1S$ state.}
\label{fig1}
\end{figure}

Of course, (\ref{eq9}) represents a non-relativistic approximation and
should not be used indiscriminately for large $p$.  However, it is
adequate for the analysis presented here.  In fig.~\ref{fig1}, we
display the bound state momentum probability distribution
corresponding to (\ref{eq9}).

Muonium bound state modifications of the muon decay spectrum and
lifetime must vanish as $\alpha$ or $m_e/m_\mu\to 0$.  It is,
therefore, useful to organize such corrections as an $\alpha^n
(m_e/m_\mu)^m$, $n,m=1,2\ldots$ expansion in the small parameters
$\alpha\simeq 1/137$ and $m_e/m_\mu \simeq 1/207$.  Terms of order
$n+m>4$ are completely negligible and can be safely ignored. 

\begin{figure} 
\begin{minipage}{16.cm}
\[
\hspace*{-15mm}
\mbox{ 
\psfig{figure=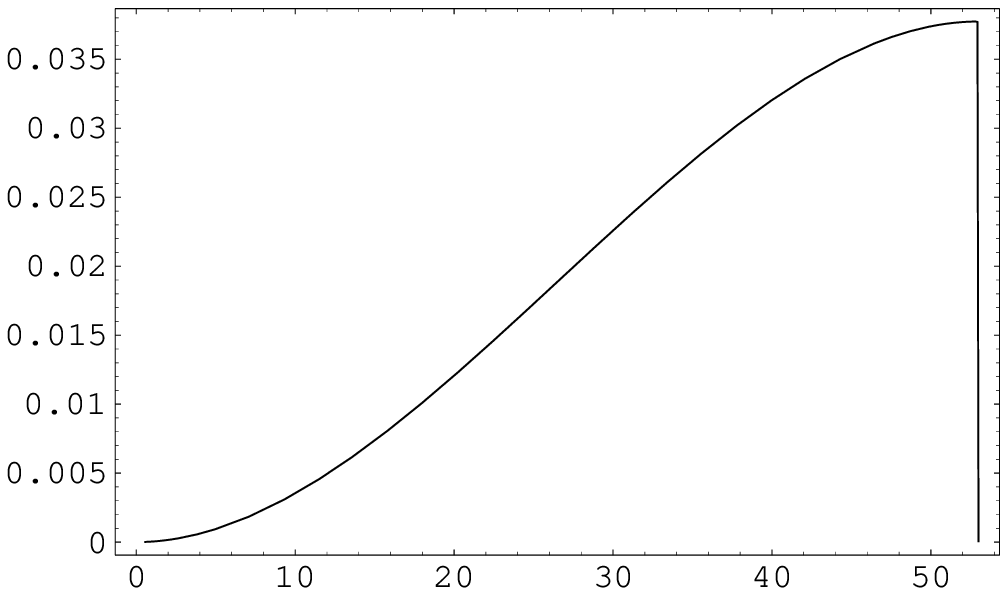,width=100mm}
}
\]
\end{minipage}
\begin{picture}(0,0)(0,0)
\put(380,15) {$E_{e^+}$ [MeV]}
\put(50,170) {${1\over \Gamma}{{\rm d}\Gamma \over {\rm d}E_{e^+}}$}
\end{picture}

\caption{Positron energy spectrum for free $\mu^+$
decay at rest.}
\label{fig:eplus}
\end{figure}

The leading ${\cal O}(\alpha m_e/m_\mu)$ effects result from the
muon's non-zero velocity in the muonium rest frame.  (For this
discussion we take the muonium atom to be at rest in the lab frame.)
The ground state velocity, $\beta_\mu \sim \alpha m_e/m_\mu$, will
modify the $e^+$ decay spectrum relative to its shape for free $\mu^+$
decay at rest (see fig.~\ref{fig:eplus}).  For a given $\beta_\mu$, it
gives rise to Doppler 
broadening which corresponds to an overall dilation along with
smearing of the positron energy over the range $(E_{e^+} \pm \beta_\mu
p_{e^+})/\sqrt{1-\beta_\mu^2}$. That distortion amounts to about $\pm
1.9$ keV near the free decay endpoint $E_{e^+} =
(m_\mu^2+m_e^2)/2m_\mu$ for $\beta_\mu \sim \alpha m_e/m_\mu$. 
Such an effect leads to a small spectral tail
beyond the usual endpoint. Of course, that enhancement is accompanied
by a depletion below the usual endpoint.  In fact, smearing corresponds
primarily to a redistribution of the spectrum.  Lorentz invariance
requires that such effects cancel in the total integrated decay rate
up to an overall time dilation factor $\gamma =
1/\sqrt{1-\beta_\mu^2}$ which slightly reduces the decay rate by
$1/\gamma$.  Averaging over the momentum distribution in
Eq.~(\ref{eq9}) or its relativistic generalization replaces
$\beta_\mu^2$ by $\langle \beta_\mu ^2 \rangle \simeq \alpha^2
m_e^2/m_\mu^2$ and one finds \cite{meta}
\begin{eqnarray}
\Gamma(M\to e^-e^+ \nu_e \bar \nu_\mu) &=& \left( 1-{\langle \beta_\mu
^2 \rangle \over 2} \right) \Gamma ( \mu^+ \to e^+ \nu_e \bar \nu_\mu),
\label{new10} \\
\tau_M &=& \tau_\mu \left( 1+{\alpha^2 m_e^2\over 2m_\mu^2}\right),
\label{new11} \\
{\tau_M - \tau_\mu\over \tau_\mu} &=& {\alpha^2 m_e^2\over 2m_\mu^2}
\simeq 6\times 10^{-10}.
\label{new12}
\end{eqnarray}
Although the bound state velocity effects on the spectrum are
relatively large
${\cal O} (\alpha m_e/m_\mu) \simeq 35$ppm  compared to the $\pm
1$ ppm experimental lifetime goal, the overall shift in
Eq.~(\ref{new12}) is completely negligible.  So, lifetime counting
experiments  that are independent of spectrum shape details are
very insensitive to small muon bound state velocity effects.

In the case of muonium stopped in matter, the bound state and thermal muonium
velocities should be approximately added in quadrature
\begin{eqnarray}
\left( \beta_\mu^{\rm total} \right)^2 
 \simeq\left( \beta_\mu^M\right)^2 +\left( \beta_\mu^{\rm thermal}\right)^2 .
\label{new13}
\end{eqnarray}
At room temperature, $\beta_\mu^{\rm thermal} \simeq 0.8 \beta_\mu^M$ and the
total time dilation shift in Eq.~(\ref{new12}) increases by about a
factor of 1.6. Nevertheless, it remains negligible at the 1 ppm level.  

The muonium bound state Coulomb potential, $\langle V\rangle \sim
-\alpha^2 m_e$, also modifies the positron decay spectrum.  Those
${\cal O}(\alpha^2 m_e/m_\mu)$ corrections correspond to $\sim -27$ eV
modifications of the spectrum.  If they were to also affect the muon
lifetime at that order, the effect would be larger than time dilation
and possibly near the $\pm 1$ ppm goal of future experiments.
 However, as we show in the Appendices, the leading effects
cancel, leaving behind contributions that are much smaller than~1\,ppm.

Aside from Doppler broadening, the bound state Coulomb interaction has
two main effects on muon decay: phase-space suppression, because the muon is
off mass shell, and final-state $e^+e^-$ interactions after the
decay. In fact these two effects largely cancel, because of charge
conservation. We present a rigorous proof of this cancelation in the
Appendices, but the nature of the cancelation is easily understood if
we replace muonium by a simpler system, where the muon is bound to a
static charge distribution with density $\rho(r) \equiv e|\psi(r)|^2$,
rather than to an electron.

Initially the muon in our simplified system is at the center of
the charge distribution with negligible kinetic energy and potential
energy $-V_0$, where $V_0=27$\,eV.  Since the muon is below mass
shell, the maximum energy allowed to the positron after
the muon decays is lowered by 27\,eV from what it is for free-muon
decay. This phase-space contraction tends to lower the decay rate.
The positron, however, is
formed at the center of our ``muonium'' atom and, because it has the same
charge as the muon, it initially has the same potential energy,
$-V_0$, by virtue of its interactions with the charge
distribution~$\rho(r)$.
The positron rapidly leaves the atom, but its asymptotic kinetic energy
is reduced from its initial kinetic energy by the 27\,eV needed for it
to climb out of the potential well due to~$\rho(r)$.
Thus the entire positron energy distribution is shifted down by
$V_0 = 27$\,eV:
\begin{equation}
\frac{\drm\Gamma_M(E)}{\drm E_{e^+}}
= \frac{\drm\Gamma_\mu(E+V_0)}{\drm E_{e^+}}
\end{equation}
This is the effect of final-state interactions. The two
effects, phase-space contraction and final-state interactions, cancel
in the total decay rate because the energy distribution and the
kinematic limit are shifted by the same amount, leading to the
same total integral for the decay rate:
\begin{eqnarray}
\int^{m_\mu/2 - V_0} \frac{\drm\Gamma_M(E)}{\drm E_{e^+}} \,\,\drm E
&=& \int^{m_\mu/2 - V_0}
\frac{d\Gamma_{\mu}(E+V_0)}{\drm E_{e^+}}\,\, \drm E
\nonumber\\
&=& \int^{m_\mu/2}
\frac{\drm\Gamma_{\mu}(E^\prime)}{\drm E_{e^+}}\,\,  \drm E^\prime
\end{eqnarray}
At very low-energies, the positron can bind to the charge
distribution and our analysis must be modified. However any such
modification is of order $\alpha^3m_e^3/m_\mu^3\ll
\alpha^2m_e^2/m_\mu^2$ (the time dilation effect) for the total decay
rate since the spectrum is highly suppressed, by phase space $\sim
pE\drm E$, in that low-energy region (see fig.~2). Such
corrections are negligible.

Other effects, e.g.~hyperfine interactions, $e^+e^-$ hard scattering
etc., are also of higher order in $\alpha$ and/or $m_e/m_\mu$ and therefore
suppressed.  Thus, overall, the time dilation effects in
(\ref{new10})--(\ref{new12}) represent the dominant bound state
correction to the muonium lifetime.


The cancelation of ${\cal O}(\alpha^2 m_e/m_\mu)$ effects in the total
$\mu^+$ decay rate is analogous to other cancelations of a similar
nature that have been previously pointed out. 

A recent example of that phenomenon has been encountered in inclusive
$B$ decays.  The total decay rate of a meson or baryon containing a
heavy $b$ quark can be approximated by a ``free'' $b$ quark decay
rate.  If the rate is expressed in terms of a short distance $b$ quark
mass, corrections to that approximation are ${\cal O}(1/m_b^2)$ rather
than ${\cal O}(1/m_b)$.  They stem from time dilation, just as we have
found for the much simpler muonium bound state, as well as from
hyperfine interactions (which for muonium are suppressed by
$\alpha^4$).  In that case, the absence of $1/m_b$ corrections is due
to color conservation
\cite{Bigi:1992su}.  Our NRQED analysis in
the Appendix is closely related to the standard $B$ analysis.

Another example of the cancelation of $V$ dependence in total decay
rates was found by H. \"Uberall \cite{ueberall60} in his classic study
of muonic atoms, $\mu^- -$Nuclei bound states.  There, the $\mu^-$ is
bound to the nucleus by a much stronger $V\sim -Z^2\alpha^2 m_\mu$ than
muonium.  Nevertheless, \"Uberall found that the Coulomb potential
reduction of initial state phase space was canceled by final state
$e^- -$Nuclei Coulombic interactions in the total decay rate.  In
fact, he also found for that example that the leading bound state
effect was time dilation which reduces the effective $\mu^-$ decay
rate in muonic atoms by
\begin{eqnarray}
\gamma^{-1} = \sqrt{1-Z^2\alpha^2} \simeq 1-{1\over 2} Z^2\alpha^2,
\qquad (Z^2\alpha^2 \ll 1).
\label{new14}
\end{eqnarray}
Of course, besides ordinary decay $\mu^- \to e^- \bar \nu_e \nu_\mu$,
the muonic atom can undergo weak capture, $\mu^-p \to \nu_\mu n$,
with protons in the nucleus.   In fact, for heavy nuclei with 
$Z {\lower.7ex\hbox{$ \;\stackrel{\textstyle>}{\sim}\;$}} 12$ 
the capture process dominates over ordinary decay.
Capture rates are generally obtained by measuring the $\mu^-$ lifetime
in a material and subtracting out the ordinary decay rate ($\mu^-\to
e^-\bar \nu_e \nu_\mu$) using the correction in (\ref{new14}).  That
prescription needs some adjustment for high $Z$ nuclei where residual
low energy $e^--$Nuclei interactions can significantly suppress the
decay rate and become comparable to time dilation \cite{huff61}.

The analog of $\mu^-p$ capture for muonium is
annihilation  $M\to \nu_e \bar \nu_\mu$.  We
have computed that rate for the $n=1$ ground state and find
\cite{Our,Muonium}
\begin{eqnarray}
\Gamma(M\to \nu_e \bar \nu_\mu) &=& 48\pi\left({\alpha m_e\over
m_\mu}\right)^3 \Gamma(\mu^+\to e^+  \nu_e \bar \nu_\mu )
\nonumber \\
&\simeq & 6.6 \times 10^{-12} \, \Gamma(\mu^+\to {\rm all}).
\label{eq13}
\end{eqnarray}
Our result is in accord with the more general
analysis of Ref.~\cite{Li:1988xb}.  That rate is of higher order in
$\alpha m_e/m_\mu$ than the time dilation correction in (\ref{new11})
and 100 times smaller.  It can be safely neglected. 

The very small $\mu^-p\to \nu_\mu n$ capture rate for protons has been
determined by comparing measurements of $\tau_{\mu^+}$ and
$\tau_{\mu^-}$ in liquid hydrogen \cite{Bardin:1981mi}. In that way,
the induced weak pseudoscalar coupling $g_p$ of the nucleon can be
inferred.  Our general analysis along with \"Uberall's study lend
theoretical support to this method.  Increasing the precision of those
measurements by a factor of 10--20 (if possible) could provide an
interesting confrontation with theory \cite{Bernard:1998rs}.

Our analysis of the muonium bound state effect on the $\mu^+$
lifetime provides a simple way of estimating other effects of a
stopping medium.  For example, in metals the fraction of stopped
$\mu^+$ in muonium is tiny.  Instead, the thermal muon is screened by
conduction band electrons.  The density of screening electrons near
the $\mu^+$ is expected to be comparable to muonium, resulting in a
similar $V$ and average $\mu^+$ velocity.  Hence, the distortion of
the $e^+$ emission spectrum due to Doppler smearing and the final
$e^+$ potential energy should be roughly the same as muonium.  
Due to Lorentz and gauge invariance, the leading effects will continue
to cancel
in the total decay rate.  The main medium effect on the $\mu^+$
lifetime will again be time dilation due to $\beta_\mu \neq 0$ (from
thermal and electromagnetic interactions) which will be negligible
$\sim {\cal O} (10^{-9})$.  In the Appendix, we describe how such
medium effects can be analyzed using nonrelativistic QED.

A possibility that we need only briefly discuss is the effect of the
medium on radiative muon decay.  Ordinary free muon decay
bremsstrahlung, $\mu^+\to e^+\nu_e\bar\nu_\mu\gamma$ (as well as
$\gamma\gamma$ and $e^+e^-$ production), is included in the ${\cal
O}(\alpha)$ and ${\cal O}(\alpha^2)$ terms of Eq.~(\ref{eq3}).  Those
effects are rather small, despite the fact that soft and collinear
bremsstrahlung can significantly modify the spectrum.  As in our above
analysis, those large logarithmic infrared effects primarily give rise
to a redistribution of the spectrum but tend to cancel in the total
decay rate \cite{kln1,kln2}.

Our general arguments carry over to radiative muonium decay.
The radiative muonium decay 
spectrum will be distorted by Doppler smearing and $V$
dependence, but those effects will largely cancel in the total decay
rate, again leaving time dilation as the primary correction.  A
similar argument can be applied to alternative medium potential
effects such as screening in metals.  

Other long distance properties of a stopping medium may be more subtle
to analyze. For example, after the $\mu^+$ decay process, an outgoing
$e^+$ might emit outer bremsstrahlung or \v{C}erenkov radiation due to
its material environment.  Those effects will modify the $e^+$ and
electromagnetic emission spectra.  However, such changes again
correspond to a redistribution of events and should cancel (to a very
good approximation) in the total decay rate or lifetime (assuming
Lorentz and gauge invariance are maintained in the analysis).  The
space-time scale associated with the actual decay process (not the
lifetime) is $1/m_\mu$.  That space-time interval is much too small
to be affected by long distance environmental conditions.

In summary we have found that lifetimes of muonium and a free $\mu^+$
differ in leading order only by time dilation effects which are $\sim
10^{-9}$ and completely negligible for future $\pm 1$ ppm
experiments. The smallness of that difference follows from Lorentz and
gauge  invariance which guarantee cancelations among considerably
larger spectrum distortions. A similar suppression of other medium
effects on the $\mu^+$ lifetime is also expected.

\section*{Acknowledgments}
We thank D. Hertzog, R. Hill, P. Kammel, T. Kinoshita, and A. Sirlin
for stimulating discussions that rekindled our interest in this
problem.  A.C. thanks N.~Uraltsev and A.~Yelkhovsky for helpful
correspondence.  This work was supported by the DOE contract
DE-AC02-98CH10886, and by a grant from the National Science
Foundation.

\appendix
\newcommand{\vve}{{e^{\tiny +}\nu\bar{\nu}}}
\newcommand{\be}{\begin{equation}}
\newcommand{\ee}{\end{equation}}
\newcommand{\bea}{\begin{eqnarray}}
\newcommand{\eea}{\end{eqnarray}}
\newcommand{\nl}{\\ \nonumber}
\newcommand{\pe}{p_e}
\newcommand{\pmu}{p_\mu}
\renewcommand{\Im}{{\mathrm Im}}
\newcommand{\psib}{\overline{\psi}}
\newcommand{\nrqed}{{\mathrm NRQED}}
\newcommand{\Dv}{{\bf D}}
\newcommand{\Ev}{{\bf  E}}
\newcommand{\Bv}{{\bf B}}

\section{Bethe-Salpeter and NRQED Analyzes}
In the text, we described how spectral shifts occur in muonium decay
but largely cancel in the total decay rate, leaving time dilation as
the primary bound state correction.  Here, a proof of those
cancelations is given using a Bethe-Salpeter bound state approach.  A
second formalism, nonrelativistic QED (NRQED) \cite{nrqed}, 
is then employed to
illustrate the important role of electromagnetic gauge invariance or
charge conservation in the cancelation.  That method provides a
powerful means of parameterizing other more general medium effects and
showing that they also lead to negligible corrections to the muon
lifetime in matter. 

\subsection{Bethe-Salpeter Analysis}
\begin{figure} 
\hspace*{35mm}\psfig{figure=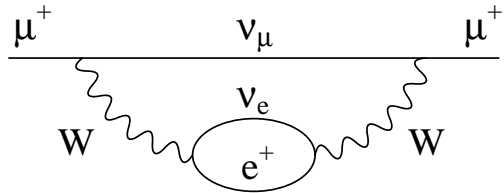,width=70mm}
\vspace*{7mm}
\caption{Self energy of the muon.  Its imaginary part, computed in the
external Coulombic field, determines the lowest order muonium width.}
\label{fig3}
\end{figure}

\begin{figure} 
\hspace*{1mm}
\psfig{figure=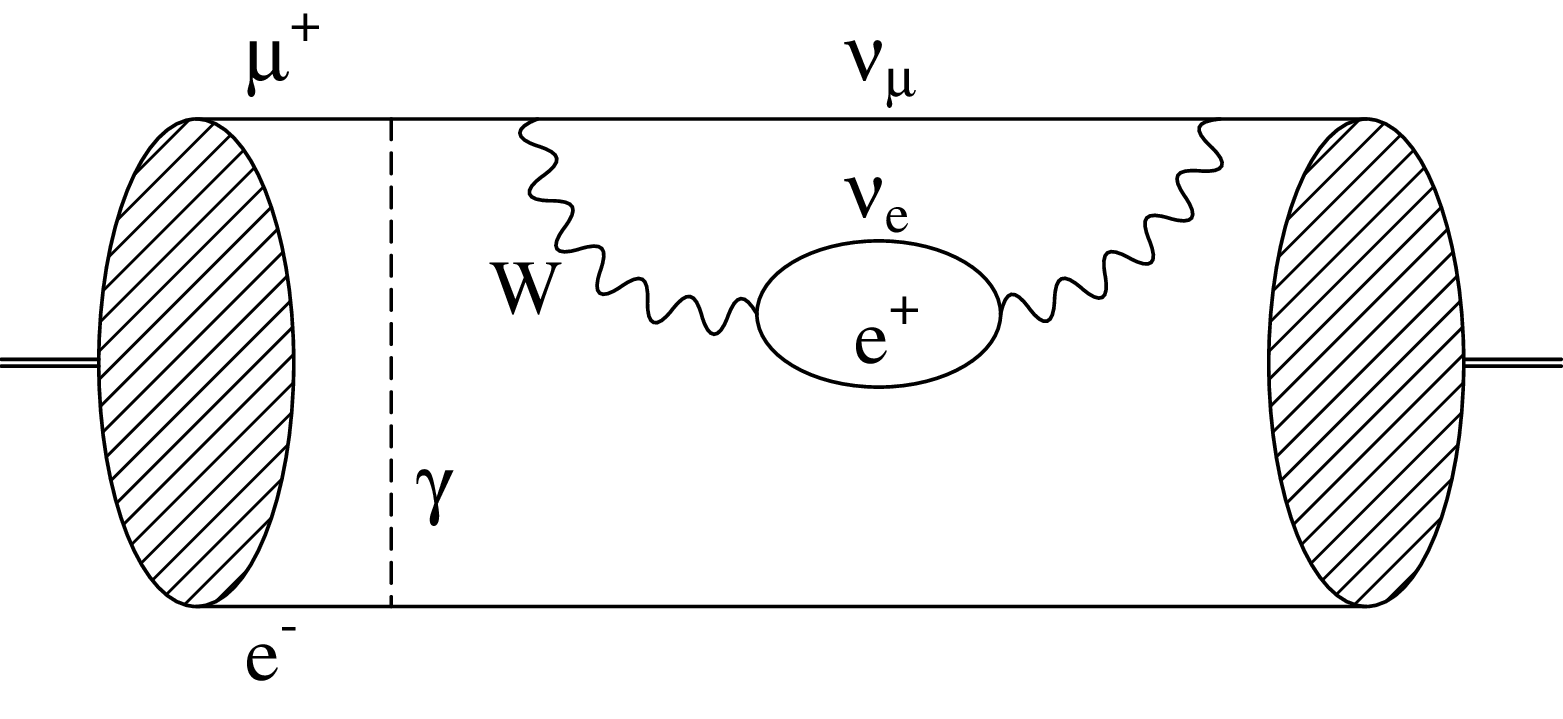,width=70mm}
\hspace*{10mm}
\psfig{figure=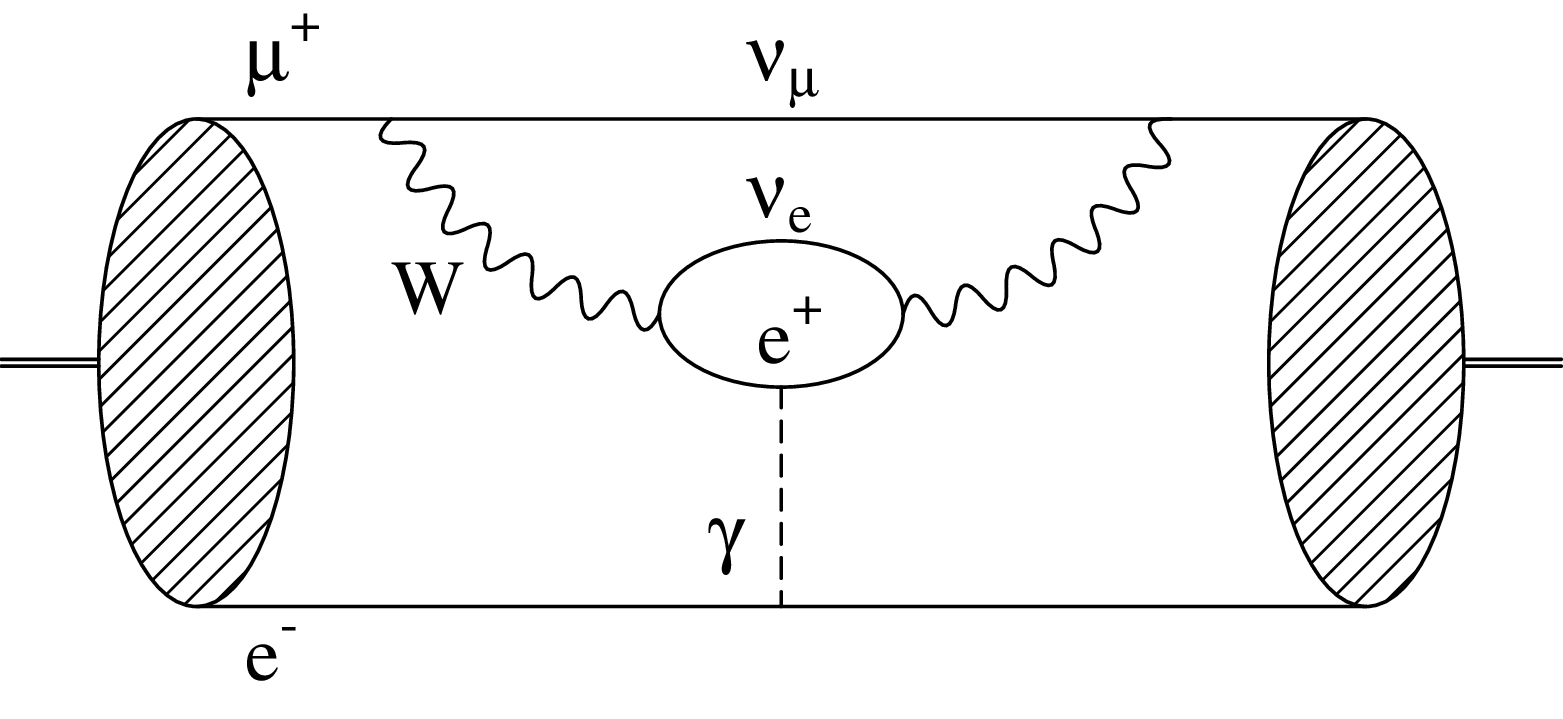,width=70mm}

\hspace*{34mm}(a) \hspace*{75mm} (b)

\vspace*{7mm} 

\caption{Coulomb interactions in muonium decay.}
\label{fig4}
\end{figure}

We can compute the leading bound state corrections to muonium decay
using standard Bethe-Salpeter perturbation theory for the bound state
energies \cite{bethe77}.  The decay rate is obtained from the
imaginary part of the 
energy ($\Gamma = -2\Im\,E$). The leading effect comes from the
$\mu^+\to\vve\to\mu^+$ contribution $\Sigma_\vve(p)$ to the muon's self
energy (see figs.~\ref{fig3} and \ref{fig4}(a)). 
Since the decay products are typically highly relativistic, we
can Taylor expand the self energy about the mass-shell momentum in the
usual fashion \cite{expa}:
\be
\Im\,\Sigma_\vve(p) = -\frac{\Gamma_\mu}{2} 
+ (p\cdot\gamma - m_\mu)\,\delta Z_\vve + \cdots
\ee
where $\Gamma_\mu$ is the free muon 
decay rate and $m_\mu$ is the muon pole mass.
In muonium, the second term in this expansion gives the leading
decay rate correction due to the bound state reduction
of the $\vve$ phase space.
The shift in the bound state 
energy due to this self energy is, in first order
perturbation theory,
\bea
\psib\,(\pe\cdot\gamma - m_e)\,\Im\,\Sigma_\vve(\pmu)\,\psi
&=& -\,\frac{\Gamma_\mu}{2}\,\,\psib(\pe\cdot\gamma - m_e)\psi
\nonumber \\
&& + \,\delta Z_\vve\,\,\psib (\pe\cdot\gamma -
m_e)(\pmu\cdot\gamma_{(\mu)}-m_\mu)\psi,
\label{sigmaval}
\eea 
where $\psi$ is the $\mu^+e^-$ wave function,
$\gamma_{(\mu)}$ denotes Dirac matrices for the muon spinor part of $\psi$,
$\pe$ and $\pmu$ are the electron and muon momenta respectively,
and integration over the wave function's momenta distribution is
implicit. The constant $\delta Z_\vve$ is of order $\Gamma_\mu/m_\mu$.

The coefficient of $-\Gamma_\mu/2$ in this expectation value is nearly
equal
to unity since the wave functions are normalized such that \cite{rhs}
\be
\psib\, \gamma^0_{(\mu)}\, (\pe\cdot\gamma - m_e)\, \psi = 1.
\ee
The standard nonrelativistic expansion
\be
\frac{\overline{u}(p) 1 u(p)}{\overline{u}(p) \gamma^0 u(p)}
\approx  1 - \frac{p^2}{2m^2}
\ee
implies, therefore, that
\be
\psib\,(\pe\cdot\gamma - m_e)\,\psi \approx \langle 1 - \beta_\mu^2/2
\rangle
\ee
where $\beta_\mu$ is the muon velocity. This factor decreases the decay
rate
for muonium; it is the time dilation caused by the muon's motion
within the atom.

The second term in expectation value~(\ref{sigmaval}) is simplified by
noting that
\be
(\pe\cdot\gamma - m_e)(\pmu\cdot\gamma_{(\mu)} - m_\mu)\,\psi = V\,\psi,
\ee
from the Bethe-Salpeter equation, where $V$ is the binding
potential (dominated by the Coulomb interaction). Thus, through
order~$\alpha^2$,
\be \label{sigmafinal}
\psib\,(\pe\cdot\gamma - m_e)\,\Im\,\Sigma_\vve(\pmu)\,\psi
\,\,\approx\,\, -\frac{\Gamma_\mu}{2}\,\langle 1 - \beta_\mu^2/2\rangle
 + \delta Z_\vve\,\,\langle V \rangle,
\ee
where the first correction to the free decay rate
is due to time dilation, and
the second correction is due to the
contraction of final-state phase space. These two corrections are of
relative orders $\alpha^2 m_e^2/m_\mu^2$ and $\alpha^2 m_e/m_\mu$
respectively.

A second contribution is relevant in order~$\alpha^2$. This is from
the imaginary part of the muon-photon vertex
correction~$\Gamma_\vve^\rho$ due to the fluctuation
$\mu^+\to\vve\to\mu^+$ --- that is, the correction obtained by attaching
a photon to the $e^+$ in the muon self-energy diagram for
$\Sigma_\vve$ (see fig.~\ref{fig4}(b)). Physically this corresponds to a
final-state interaction between the outgoing positron (from the decay)
and the electron (from the atom).  The momentum transfer carried by
atomic
photons ($\approx \alpha m_e$) is tiny compared to the typical $\vve$
momenta, and so we can Taylor expand the vertex function to obtain the
leading contribution:
\be \label{gamma-exp}
\Im\,\Gamma_\vve^\rho = -\delta Z_\vve\,\gamma^\rho + \cdots,
\ee
where the remaining terms on the right all vanish on mass shell for
zero momentum transfer. The constant $\delta Z_\vve$ here is the same
as that appearing in the self energy; this is required by QED gauge
invariance and can be proven using Ward identities in the standard
fashion. This vertex correction results in a perturbation given by
\be
-\delta Z_\vve\,\psib V \psi\,\,=\,\, -\delta Z_\vve\,\langle V
 \rangle
\ee
through order $\alpha^4$ corrections (due to higher order terms in the
expansion~(\ref{gamma-exp}) of the vertex correction).
Remarkably, this final-state interaction completely cancels the
part of the self-energy correction due to the contraction of phase
space. Thus, through order~$\alpha^2m_e^2/m_\mu^2$ the muonium decay rate is
\be
\Gamma_{\mu e} = \Gamma_\mu \,\langle 1 - \beta_\mu^2/2\rangle .
\ee
It should be emphasized that the cancelation of final-state and
phase-space effects is a particular consequence of QED gauge
invariance. Had photons been massless scalars instead of vector gauge
bosons, for example, the final-state interaction would have vanished
and the contraction of phase space would have been the dominant
decay rate correction.

\subsection{NRQED and Muon Decay}
The Bethe-Salpeter formalism is awkward for analyzing muonium or other
medium effects on muon
decay with greater precision. A better approach follows from the
observation that the
fluctuation $\mu^+\to\vve\to\mu^+$ occurs over distances, of order
$1/m_\mu$, which are tiny compared with atomic length scales. Thus, the
effects of such fluctuations can be modeled in a low-energy effective
Lagrangian by local, gauge-invariant interactions. A standard
nonrelativistic effective field theory used in high-precision studies
of muonium is nonrelativistic QED (NRQED). The corrections due to muon
decay, being local, are easily included in that formalism
as (nonunitary) corrections to the standard NRQED Lagrangian:
\bea
{\cal L}_\nrqed &=& \psi_\mu^\dagger \left\{
i D_t + \frac{\Dv^2}{2m_\mu} + \cdots \right. \nl
&& + \frac{i\Gamma_\mu}{2} \left( 1 + c_1 \frac{\Dv^2}{2m_\mu^2}
- c_2 \frac{\Dv^4}{8m_\mu^4} + \cdots \right. \nl
&& \quad\quad + d_1 \frac{\psi^\dagger_e \psi_e}{m_\mu^3}
+ d_2 \frac{\sigma\cdot\psi^\dagger_e\sigma\psi_e}{m_\mu^3}
+ \cdots \nl
&& \quad\quad\left.\left. + f_1 \frac{e\sigma\cdot\Bv}{m_\mu^2}
+ f_2 \frac{e\nabla\cdot\Ev}{m_\mu^3}+\cdots\right)\right\}\psi_\mu.
\eea
Here $D_t$ and $\Dv$ are the temporal and spatial gauge-covariant
derivatives,  $\Ev$ and $\Bv$ are the electric and magnetic field
operators, and $\psi_\mu$ and $\psi_e$ are two-component Pauli fields
for muons and electrons. This effective theory provides a rigorous,
systematic framework
for analyzing nonrelativistic muons, including relativistic
corrections as well as muonium and medium effects on 
muon decay, to whatever precision is desired.

The couplings $c_1$, $c_2$, $d_1$\,\ldots\,in the effective Lagrangian
are all dimensionless and have perturbative expansions in~$\alpha$.
The $c_i$'s are determined by computing muon
decay, using the effective theory, for muons in different reference
frames. They are tuned until the decay rate predicted by the effective
theory agrees with the relativistic result
\be
\Gamma_\mu(\beta_\mu) = \Gamma_\mu(0) \,\left(1 - {\beta_\mu^2\over 2} -
{\beta_\mu^4\over 8} - \cdots\right);
\ee
at tree-level $c_1=c_2=1$. The $d_i$'s incorporate the contribution
from muon-electron annihilation: $\mu e \to \nu\bar{\nu}$ (see
Eq.~(\ref{eq13})). The $f_i$'s are computed by comparing the on-shell,
renormalized vertex function $\Gamma_\vve$, expanded in powers of the
external momenta, with predictions from the effective theory.

Two features of the effective theory deserve comment. One is that
there are no terms of the form
\be
\frac{i\Gamma_\mu}{m_\mu} \,\psi^\dagger_\mu eA^0 \psi_\mu,
\ee
where $A^0$ is the scalar potential in electromagnetism (that is, the
Coulomb potential in an atom). Such a term would be required to
generate a contribution such as that due to either phase-space
contraction (second term in Eq.\,(\ref{sigmafinal})) or final state
$e^+e^-$ interactions alone. It cannot arise
here because it would violate QED gauge invariance. This confirms the
detailed Bethe-Salpeter analysis of the previous section which
demonstrated the cancelation of such terms. 

The second observation is that the parameter $\Gamma_\mu$ appearing
in the effective Lagrangian is precisely the free muon decay rate,
to all orders in $\alpha$; it is {\em not} a
``bare'' decay rate that is further renormalized by medium dependent
radiative 
corrections. This is because the operator $\psi^\dagger_\mu\psi_\mu$ in
\be
\delta{\cal L} \equiv \frac{i\Gamma_\mu}{2} \,\psi^\dagger_\mu\psi_\mu
\ee
is a conserved quantity in the nonrelativistic theory; it is the
pure number operator that counts muons and so cannot be renormalized.
Consequently
\be
\frac{\langle \phi | \delta{\cal L} | \phi \rangle}{\langle \phi |
 \phi \rangle}   = \frac{i \Gamma_\mu}{2}
\ee
for any state $|\phi\rangle$ that contains a single
muon\,---\,including muonium states as well as arbitrarily complicated
many-electron states or other medium effects 
that describe muons inside bulk materials. This
``no-renormalization theorem'' implies that the only corrections to
the free muon decay rate in such systems come from the explicit
nonunitary, factorizable correction terms in the effective Lagrangian,
the terms with couplings $c_i$, $d_i$, $f_i$\,\ldots\, that are
directly related to muon decay.  As discussed in the text, these
corrections are typically of order~1\,ppb or smaller. Other effects
not of this form, due to external photons or multi-electron
excitations, cannot modify the total decay rate.


\begin{thebibliography}{10}
\bibitem{meta}
The $2S$ state of free muonium is metastable \protect\cite{muoniu}, decaying to
  $1S+\gamma\gamma$ in about 1/7 sec. The value of $\langle \beta^2_\mu\rangle
  = \alpha^2 m_e^2/4m_\mu^2$ in that configuration leads to a factor of 4
  reduction in the corresponding time dilation corrections of
  Eqs.~(\protect\ref{new10})--(\protect\ref{new12}).
\bibitem{expa}
The expansion is analytic because the final decay state is far from any
  threshold. Bound state effects on the low energy $e^+$ spectrum modify the
  total decay rate at $\order\left( \alpha^3 m_e^3/m_\mu^3\right)$ and can,
  therefore, be ignored in this analysis.
\bibitem{rhs}
The right-hand side of the normalization condition would be the electromagnetic
  form factor of muonium at zero momentum transfer if the electron had no
  electric charge, and provided that the potential is instantaneous (which is
  true up to $\order(\alpha^3)$~corrections which we ignore). Charge
  conservation or gauge invariance demands that this form factor equal unity.
\end{thebibliography}

\end{document}